# Growth of Antiperovskite Oxide $Ca_3SnO$ Films by Pulsed Laser Deposition

Makoto Minohara[1,2]*, Ryu Yukawa[1], Miho Kitamura[1], Reiji Kumai[1,2], Youichi Murakami[1,2] and Hiroshi Kumigashira[1,2]

*Corresponding author Tel: +81 29 864 5586

E-mail: minohara@post.kek.jp

[1]*Photon Factory, Institute of Materials Structure Science (IMSS), High Energy Accelerator Research Organization (KEK), Tsukuba, Ibaraki 305-0801, Japan*

[2]*Department of Materials Structure Science, SOKENDAI (The Graduate University for Advanced Studies), Tsukuba, Ibaraki 305-0801, Japan*

**Abstract**

We report the epitaxial growth of $Ca_3SnO$ antiperovskite oxide films on (001)-oriented cubic yttria-stabilized zirconia (YSZ) substrates by using a conventional pulsed laser deposition (PLD) technique.  In this work, a sintered $Ca_3SnO$ pellet is used as the ablation target.  X-ray diffraction measurements demonstrate the (001) growth of $Ca_3SnO$ films with the antiperovskite structure and a cube-on-cube orientation relationship to the YSZ substrate.  The successful synthesis of the antiperovskite phase is further confirmed by x-ray photoemission spectroscopy.  These results strongly suggest that antiperovskite-oxide films can be directly grown on substrates from the target material using a PLD technique.

**Keywords**

A3. Pulsed laser deposition

B1. Antiperovskite oxides

**Main text**

**1. Introduction**

Perovskite oxides are an important class of materials that exhibit novel functional properties [1].  Recently, their counterpart materials, namely antiperovskite oxides $A_3B$O ($A$ = Ca, Sr, Ba; $B$ = Sn, Pb), have been attracting attention as an alternative platform for prospecting unique physical properties.  Following the theoretical prediction that some materials in the antiperovskite-oxide family have three-dimensional Dirac fermions [2], a number of experimental and theoretical efforts have been devoted to exploring the possible presence of bulk Dirac fermions in antiperovskite oxides [3-8].  Since this unique electronic structure is predicted to appear in the bulk and not at the surface, heterostructures composed of perovskite and antiperovskite oxides might also be promising candidates for exploratory studies of novel physical properties, analogous to the studies of perovskite oxide heterostructures [9-13].  Despite these intriguing research opportunities, there are few reports on the growth of antiperovskite-oxide films [7,8].

Since the pulsed laser deposition (PLD) technique is capable of growing films from a wide variety of materials, this technique is well-suited for the rapid investigation of antiperovskite-oxide films. However, since the use of PLD for the growth of these films is still in its infancy, whether antiperovskite-oxide films can be really grown by PLD remains to be convincingly answered. X-ray diffraction (XRD) measurements from the reported antiperovskite-oxide films grown by an "unconventional" PLD technique, where the films were grown by alternating deposition of $A$O and $B$O$_2$ targets, did not show any of the characteristic Bragg reflections from (00$l$) planes with an odd $l$ that should be expected for an antiperovskite phase [7]. Therefore, the development of reliable methods for the growth of antiperovskite-oxide films by PLD is necessary, not only for exploring novel functionalities, but also for elucidating unique band structures via spectroscopic measurements [14-16].

In this study, we report the conventional PLD growth of an antiperovskite-oxide Ca$_3$SnO film by using the antiperovskite-oxide ceramic target as the source material. By varying the substrate temperature ($T_{\text{sub.}}$) and the oxygen partial pressure ($P_{\text{O2}}$) during growth, we identify the optimum growth condition for Ca$_3$SnO thin films. XRD results indicate the formation of the antiperovskite structure and its cube-on-cube epitaxial relationship to the substrate. The successful growth of antiperovskite phase is further confirmed by x-ray photoemission spectroscopy (XPS): The shape of the valence band spectrum is qualitatively reproduced by the first-principles calculation for Ca$_3$SnO.

**2. Experimental Section**

Ca$_3$SnO films were grown on (001)-oriented yttria-stabilized zirconia (YSZ) single-crystal substrates in a vacuum chamber by a PLD technique using a sintered $Ca_3SnO$ target. A Nd-doped yttrium aluminum garnet laser was used for ablation in its frequency-tripled mode ($\lambda$ = 355 nm) at a repetition rate of 1 Hz. Prior to film deposition, YSZ substrates were annealed in air, in an electric furnace at 1400 ºC, to obtain an atomically flat surface. To optimize the growth conditions, $T_{sub.}$ and $P_{O2}$ were varied under a fixed laser fluence of ~4.6 mJ/cm$^2$. The surface crystalline quality during growth was monitored *in situ* by reflective high-energy electron diffraction (RHEED).

The epitaxial relationship and crystalline quality were analyzed using *ex situ* synchrotron-based XRD, which was carried out at beamline 7C in the Photon Factory, KEK. The energy of the incident x-ray was 9 keV. For XRD measurements, Au was deposited as a capping layer to prevent degradation by moisture in the air. XPS measurements were performed using a VG-Scienta R3000 electron energy analyzer with a monochromatized Al $K\alpha$ x-ray source ($h\nu$ = 1486.6 eV). For XPS measurements, the samples were transported to an analysis chamber without air exposure using a home-built vacuum suitcase. Binding energies were calibrated by measuring a gold film, electrically connected to the samples. All spectra were acquired at room temperature with a total energy resolution of 500 meV.

First-principles calculations based on density-functional theory were carried out in the framework of the Perdew-Burke-Ernzerhof-type generalized-gradient approximation [17] using the WIEN2k code [18], where spin-orbit interactions are taken into account. An idealized cubic structure was assumed with the empirical lattice constant of $a$ = 4.83 Å [19]. The corresponding Brillouin zone was sampled by 20 × 20 × 20 momentum points.

## 3. Results and discussion

### 3.1 Optimization of the growth conditions for Ca$_3$SnO films

In general, the guidance of Ellingham diagrams, in which the stable valence states of simple metal oxides are mapped in terms of standard-state Gibbs free energy and temperature, is useful in deducing optimal growth conditions. However, since Ca$_3$SnO exhibits an unusual Sn valence state of Sn$^{4-}$, it is difficult to identify the processing "sweet spot" using the Ellingham diagram, as there are no thermodynamic data available for Sn$^{4-}$ [20,21]. Therefore, we optimize Ca$_3$SnO thin film growth by varying $T_{sub.}$ and $P_{O2}$ over a rather wide range of values. Figure 1(a) shows the RHEED patterns for the thin films grown at various $T_{sub.}$ and $P_{O2}$. There are two discrete regions showing the clear streak patterns indicative of the epitaxial growth of certain oxides; A region at $T_{sub.}$ = 900 ºC without oxygen flow and another between $T_{sub.}$ = 700 and 750 ºC under a $P_{O2}$ of $10^{-6}$ to $10^{-7}$ Torr. Since the RHEED pattern changes in a discontinuous manner, we surmise that different oxides are grown under these two discrete conditions. In order to survey the chemical composition of these thin films, we performed auger electron spectroscopy (AES). Figure 1(b) shows AES results for samples α ($T_{sub.}$ = 900 ºC, without oxygen flow) and β ($T_{sub.}$ = 700 ºC, $P_{O2}$ = $10^{-7}$ Torr), representative of the above-mentioned regions [also see Fig. 1(a)]. The AES results reveal the difference between these samples. While sample α does not contain Sn ions, sample β does. The disappearance of Sn for the sample α may originate from the high volatility of Sn and its oxides (SnO and SnO$_2$) [22] at $T_{sub.}$ = 900 ºC, strongly suggesting that only the growth of the remnant CaO$_x$ is stabilized at this higher temperature region. From these results, the optimum conditions for Ca$_3$SnO growth could be in the range of $T_{sub.}$ = 700 to 750 ºC under a $P_{O2}$ of $10^{-6}$ to $10^{-7}$ Torr.

3.2 Structural analysis

We performed XRD measurements to verify the formation of antiperovskite $Ca_3SnO$. Figure 2(a) shows the out-of-plane XRD $2\theta$-$\theta$ pattern for the sample β. Some characteristic peaks are clearly observed. In order to index these peaks, we compare the measured $2\theta$-$\theta$ XRD pattern with the calculated patterns of $Ca_3SnO$ (antiperovskite structure), $CaSnO_3$ (perovskite structure), and $CaO$ (rocksalt structure), since $CaSnO_3$ and $CaO$ are also potentially synthesized under such an oxidizing atmosphere [20-22]. The calculated (00$l$) Bragg peaks of the three potential oxides are shown in the Fig. 2(b). Comparison of the measured XRD pattern with the calculated patterns shows that the observed peaks in Fig. 2(a) correspond to the characteristic (00$l$) Bragg peaks of $Ca_3SnO$. Since rocksalt $CaO$ and $Ca_3SnO$ have near identical lattice constants, a (00$l$) peak with an odd $l$ is key to distinguishing these two structures. Since a (001) peak is observed, the XRD results provide strong evidence for the formation of antiperovskite $Ca_3SnO$ films by PLD under the identified optimum conditions, although the existence of some impurity phases is surveyed. Considering both the formation energy in the materials project database [23] and the calculated peak positions of candidate materials comprised of Ca, Sn and O atoms [23], the impurity phases might be $Ca_2Sn_3O_8$ or $Ca_2SnO_4$. Although the precise identification of the impurity phase is difficult at the moment, the existence of the impurity phase implies an off-stoichiometry composition in the $Ca_3SnO$ film. Indeed, the intensity ratio of (001) to (002) Bragg peaks is evaluated to be 0.18, which is slightly smaller than the ideal ratio of 0.27, suggesting the presence of structural disorder in the $Ca_3SnO$ films owing to the existence of impurity phases and a possible off-stoichiometry composition [8].

Next, we investigate the crystalline quality of $Ca_3SnO$ films and its epitaxial relationship with YSZ (001) substrate. As shown in Fig. 2(c), the full-width at half-maximum of the rocking curve for the (002) diffraction is evaluated to be around 0.14º. This value is one order of magnitude narrower than that of $Sr_3SnO$ antiperovskite films grown by molecular beam epitaxy [8]. This result identifies the current PLD-grown films as being highly crystalline, despite the impurity phases observed in the $2\theta$-$\theta$ pattern. Figure 2(d) shows the *phi*-scan of XRD patterns on {202} reflections for $Ca_3SnO$ films, together with that of YSZ substrates with cubic symmetry. The peaks are clearly observed every 90° for the film as well as the substrate, reflecting the four-fold symmetry of antiperovskite structure. This result indicates the cube-on-cube epitaxial relationship between the YSZ (001) substrate and the $Ca_3SnO$ film.

3.3 Chemical state analysis

In order to analyze the chemical state of constituent elements, XPS measurements were performed on the $Ca_3SnO$ film (sample β). The XPS spectra for Sn-$3d$, Ca-$2p$, and O-$1s$ core levels are shown in Figs. 3(a), (b), and (c), respectively, along with the results of peak deconvolution by curve-fitting analysis. Since the intensity of the peaks at higher binding energies became stronger with increasing photoelectron emission angle (not shown), the higher (lower) peaks can be assigned to the surface (bulk) components for all core levels. For the Sn-$3d$ core level, the bulk and surface components for Sn $3d_{5/2}$ states are evaluated to be positioned at energies of ~ 484.8 eV and ~486.7 eV, respectively. In terms of chemical shift, the surface component can be assigned to either $Sn^{4+}$ (486.0–487.3 eV) or $Sn^{2+}$ (485.6–487 eV) [24], suggesting the segregation of Sn oxides ($SnO_2$ or $SnO$, respectively) to the surface of the $Ca_3SnO$ films.

On the other hand, the peak position of the bulk component is unusual in oxides: The energy of ~484.8 eV for Sn $3d_{5/2}$ states is close to that of Sn metal or its intermetallic alloys ($Sn^0$: 484.3–485.2 eV) [24]. The unusual chemical shift of Sn ions might imply the existence of possible $Sn^{4-}$ states in $Ca_3SnO$ as following reasons, although the chemical shift of $Sn^{4-}$ is not known yet. Assuming that the Sn ions in $Ca_3SnO$ are close to $Sn^0$ states (antiperovskite $Ca_3SnO$ is an intermetallic), the Ca ions should be also close to $Ca^0$. However, as shown in Fig. 3(b), the peak position of the bulk Ca $2p_{3/2}$ states (346.8 eV) is significantly different from that of Ca metal and its alloys (344.9-346.0 eV) [24]. Based on the chemical shifts of $Ca^{2+}$ states (346.1–347.3 eV) [24], the valence of Ca ions in the $Ca_3SnO$ film is assigned to $Ca^{2+}$. Moreover, the chemical shift of the O-$1s$ core level can be also ascribed to $O^{2-}$ as shown in Fig. 3(c) [24]. From these experimental results, it could be reasoned that Sn ions in $Ca_3SnO$ are close to an unusual valence state of $Sn^{4-}$, although further theoretical investigation is necessary to clarify such an unusual the chemical states of Sn ions in $Ca_3SnO$.

3.4 Analysis of electronic structure

The successful synthesis of the antiperovskite phase of $Ca_3SnO$ is further supported by valence band (VB) spectra shown in Fig. 4. The upper panel of Fig. 4 shows the VB spectrum of $Ca_3SnO$ film (sample β), together with the density of states (DOS) obtained by the first-principles calculations. The reported XPS spectra of CaO [25], $CaSnO_3$ [26], $SnO_2$ [27], SnO [27], and Sn metal [28] are shown in the lower panel of Fig. 4 as references. Overall, the VB spectrum of $Ca_3SnO$ film is significantly different from those of the referenced materials. The characteristic feature in the $Ca_3SnO$ spectrum is the existence of a band in the energy range between the Fermi level and ~3 eV. This feature of the VB spectra is approximately captured by the first-principles calculation for $Ca_3SnO$, although there are some discrepancies in the peak positions for bands located at higher binding energies (upper panel, Fig. 4). Therefore, the XPS results provides additional confirmation of the formation of $Ca_3SnO$.

## 4. Conclusions

We have demonstrated that the antiperovskite oxide $Ca_3SnO$ film can be grown on a YSZ (001) substrate by using a $Ca_3SnO$ target in a conventional pulsed laser deposition (PLD) technique. By varying the temperature of the substrate and the partial pressure of oxygen during growth, we identified the optimal conditions for the PLD growth of $Ca_3SnO$ films. The successful growth of epitaxial $Ca_3SnO$ films on YSZ substrates is confirmed by x-ray diffraction measurements. The *phi*-scan of {202} reflections reveals the cube-on-cube epitaxial relationship between the $Ca_3SnO$ film and the YSZ (001) substrate. The synthesis of the antiperovskite phase of $Ca_3SnO$ is further supported by x-ray photoemission spectroscopy. The results show that antiperovskite-oxide films can be directly synthesized from the target by using a PLD technique just like several other widely-studied oxide films. The present demonstration of the growth of antiperovskite oxides by PLD might be a promising avenue for the research of antiperovskite oxides in film form as well as artificial antiperovskite/perovskite oxide structures.

**Acknowledgments**

We thank Hironori Nakao for useful discussions. This work was supported by a Grant-in-Aid for Scientific Research (No. B25287095 and 16H02115) and a Grant-in-Aid for Young Scientists (No. 15K17470) from the Japan Society for the Promotion of Science (JSPS) as well as the MEXT Elements Strategy Initiative to Form Core Research Center. The work at KEK-PF was performed under the approval of the Program Advisory Committee (Proposals No. 2013S2-002 and 2015S2-005) at the Institute of Materials Structure Science, KEK.

**Figure Captions**

**Fig. 1** (Color online) (a) A map of RHEED patterns from thin films grown on a YSZ (001) substrate at various $T_{sub.}$ and $P_{O2}$. (b) Auger electron spectra of thin films labeled α and β that are grown at $T_{sub.}$ = 900 ºC with no oxygen flow (red) and $T_{sub.}$ = 700 ºC with $P_{O2}$ = $10^{-7}$ Torr (blue), respectively.

**Fig. 2** (Color online) (a) The $2\theta$-$\theta$ XRD pattern of sample β. The asterisks indicate the peaks from a capped gold film, while the diamonds are from impurities. (b) Calculated XRD patterns for $Ca_3SnO$, $CaSnO_3$, and CaO. The bars correspond to (00*l*) peaks. (c) The rocking curve for the (002) Bragg reflection of grown films. (d) *phi*-scans for {202} reflection of the YSZ substrate (upper panel) and grown thin film (lower panel).

**Fig. 3** (Color online) XPS spectra from the $Ca_3SnO$ film (sample β) for (a) Sn-3*d*, (b) Ca-2*p*, and (c) O-1*s* core levels. The fitting results are overlaid by green, blue, and red curves, which correspond to the surface component (*S*), bulk component (*B*), and their summation, respectively.

**Fig. 4** (Color online) (upper panel) The valence band spectrum of a $Ca_3SnO$ film (sample β), together with the DOS obtained from the first principle calculation. In order to emphasize the gap-like feature in the $Ca_3SnO$ film near the Fermi level ($E_F$) of the sample, the Fermi edge of gold film is superimposed. (Lower panel) The reference spectra of CaO [25], $CaSnO_3$ [26], $SnO_2$ [27], SnO [27], and Sn metal [28].

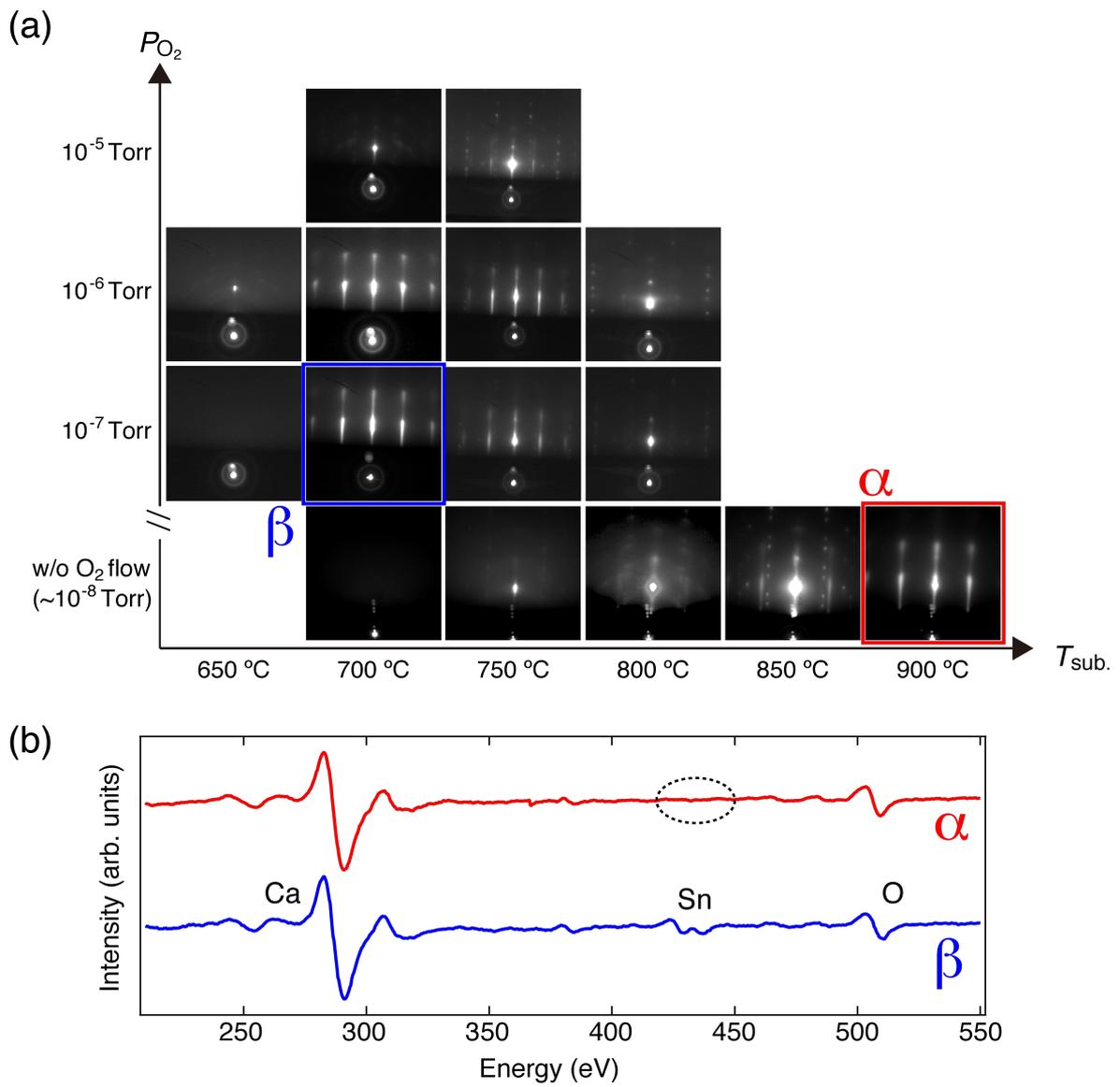

Fig. 1. (Color Online)

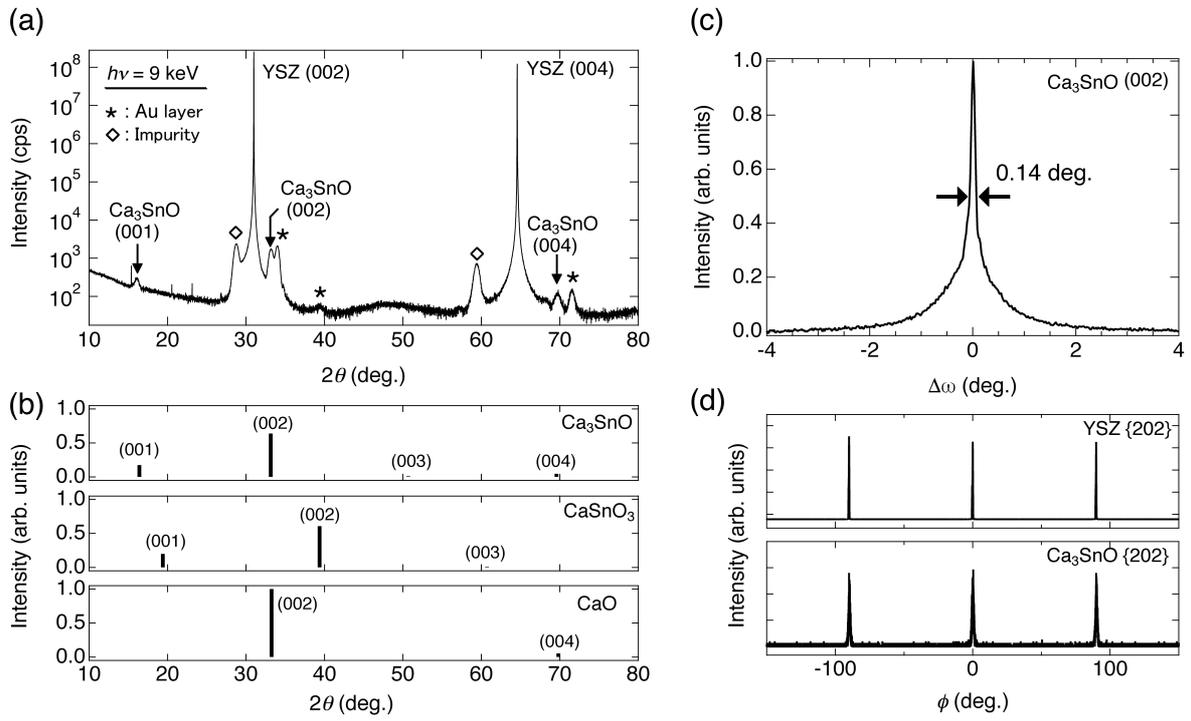

Fig. 2. (Color Online)

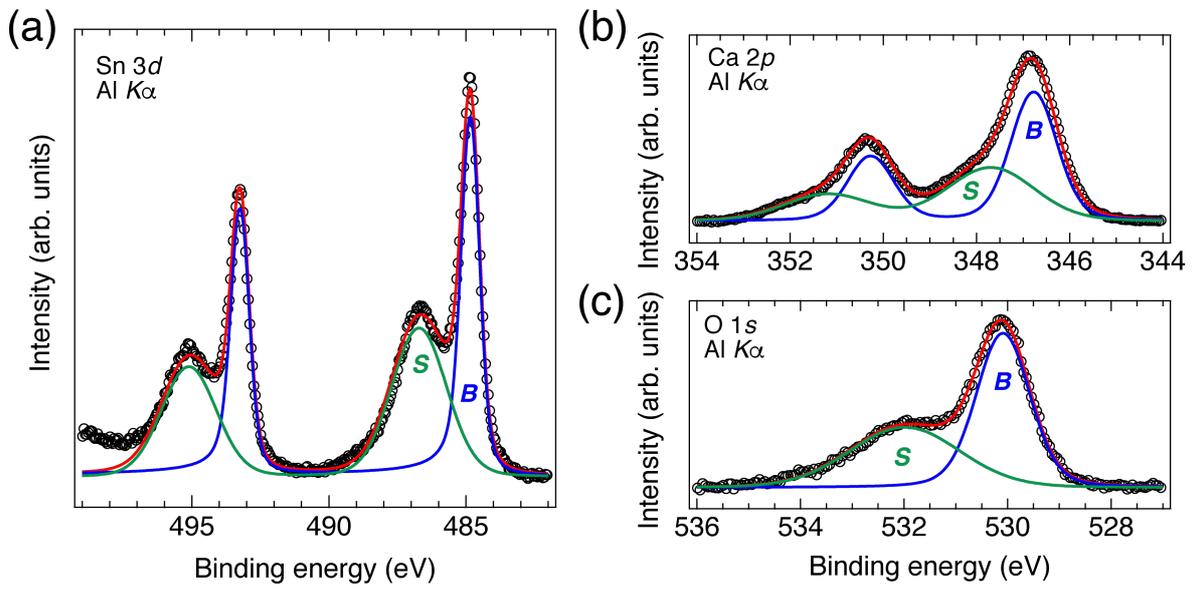

Fig. 3. (Color Online)

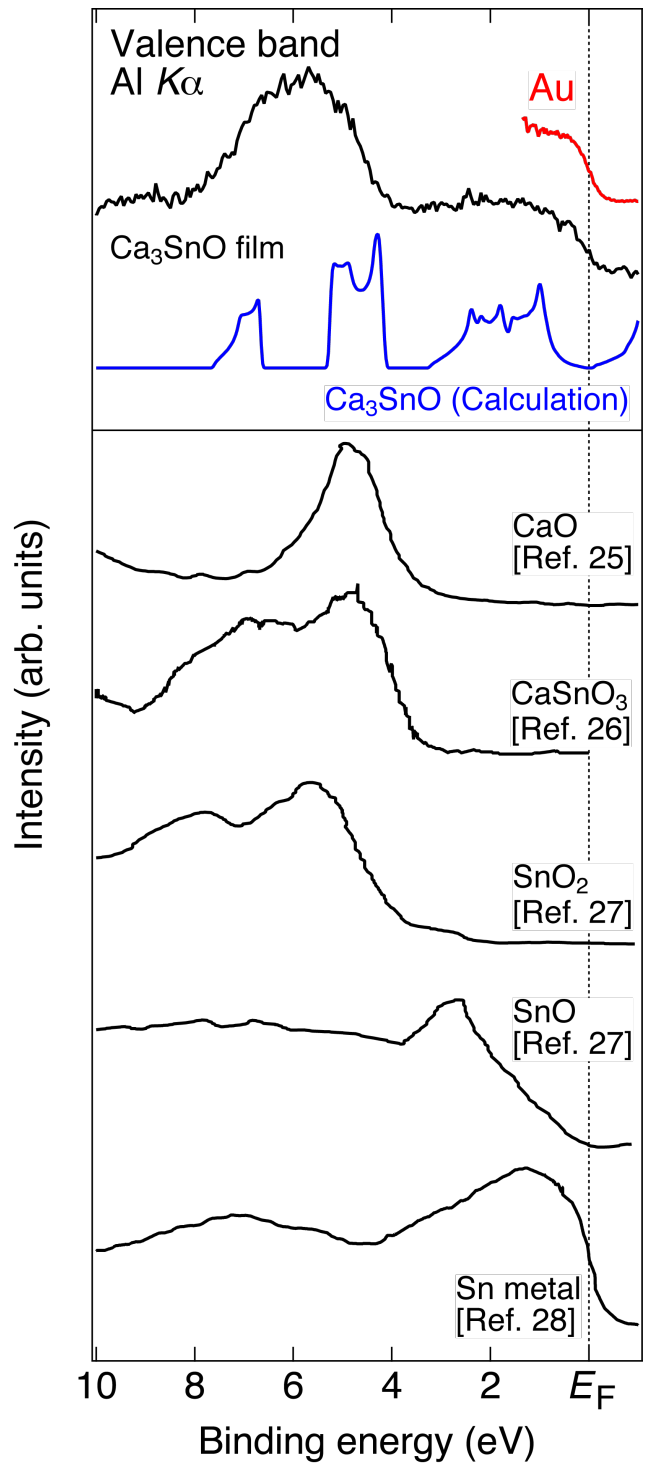

Fig. 4. (Color Online)